\documentclass[%
3p,
preprint,
]{elsarticle}

\usepackage{graphicx}
\usepackage{dcolumn}
\usepackage{bm}
\usepackage{epsfig}
\usepackage{amsfonts}
\usepackage{amsmath}
\usepackage{amssymb}
\usepackage{xcolor}
\usepackage[normalem]{ulem}

\usepackage[titletoc,toc,title]{appendix}

\usepackage{xcolor}

\begin{document}


\title{Study of  the transition from  resonance to bound states in  quantum dots embedded on a  nanowire using the $\mathbf{k}\cdot\mathbf{p}$ method}

\author{Natalia Giovenale\corref{corresp}}
\ead{ngiovenale@famaf.unc.edu.ar}
\author{Omar Osenda}%
\address{Facultad de Matem\'atica, Astronom\'{\i}a, F\'{\i}sica y Computaci\'on, Universidad Nacionald de C\'ordoba and Instituto de F\'{\i}sica Enrique Gaviola - CONICET,  Av. Medina Allende s/n, Ciudad Universitaria, CP:X5000HUA C\'ordoba, Argentina.}%
\cortext[corresp]{Corresponding author}


\date{\today}

\begin{abstract}
We study the band structure  of semiconductor nanowires with quantum dots embedded in them. The band structure is calculated using the Rayleigh-Ritz variational method. We consider quantum dots of two different types, one type is defined by electrostatic potentials applied to the nanowire, while the other one is defined by  adding materials with band offsets with respect  to the band parameters of the nanowire. We are particularly interested in the appearance of discrete energy levels in the gap between the conduction band and the valence band of the nanostructure, and in the dependence of the energy of these levels with the intensity of a magnetic field applied along the wire. It is shown that several scenarios are possible, being of particular interest the possibility of transforming states of the discrete into resonances and vice versa.

\end{abstract}


\maketitle


\section{\label{sec:introduction}Introduction}

The band structure of semiconductor nanowires has been studied intensively over the last years for several reasons. They can be manufactured in many different ways, changing the constituent materials, the characteristic lengths, coating them with other materials to change their functionality, etc. Each new configuration, or change of  the materials used, leads to novel characteristics. In this sense, the band structure can be selected to fulfill different technological needs. 

The study of the optical and electronic properties of nanowires whose crystalline structure is zincblende type, shows that the energy of the lowest state of the conduction band has a strongly non-linear dependence with the intensity of an external magnetic field, departing from the simple Landau level behavior, for instance showing Aharanov-Bohm oscillations. These oscillations can also be observed in the most energetic state of the valence band \cite{Kishore2014}. When the null-field band structure is studied, the so-called "camel-backs" also appear in the band structure \cite{Kishore2012}. By camel-back it is understood that the maximum value of the energy of the valence band can not be found at the center of the Brillouin zone, but for non-trivial values of the Bloch wave vector. A similar behavior appears on the conduction band. All these phenomena strongly depend on the radius of the nanowire and occur in simple nanowires or with core-shell structures, and they can be found for nanowires consisting of III–V semiconductor materials \cite{Kishore2014}.

The possibility of modifying the structure of nanowire bands by introducing quantum dots or applying external electrostatic potentials has also attracted much interest. The nanostructure thus defined can give rise to bound states, associated with energy levels within the gap between the valence and conduction bands, and to resonances (or metastable states) that do not exist in the unstructured nanowire. Furthermore, if time-independent and spatially constant external magnetic fields are applied to the nanostructure, it is possible to transform resonances into bound states and vice versa. Resonance binding is the term used to describe the transition from a resonance to a bound state. 

From a theoretical point of view, most of the studies that show the possibility of "binding of resonances" in this type of nanostructure have been done using the effective mass approximation (EMA).  In this approximation, the energy of the conduction band states basically behaves like Landau states, {\em i.e.} its energy increases when the strength of the magnetic field does, and for strong enough magnetic fields, the behavior is linear. This phenomenon has been studied for one- and two-electron systems \cite{Ramos2014,Garagiola2018}. 

The stabilization of metastable states due to the presence of magnetic fields 
was first analyzed in atom-like systems. Avron, Herbst and Simon, in reference \cite{Avron1977}, argued 
about the existence of negative Helium ions, a fact that was numerically tested 
very recently~\cite{Varga2014}. In atomic-like systems the field strengths 
necessary to show that the width of a resonance is zero are about $10^5$ 
Teslas \cite{Ho1997}. Having this in mind, it is natural to ask if laboratory 
attainable magnetic fields strengths can bind few-electron resonance states in 
nano-devices. As early as in 1989, Sikorski {\em et al.} \cite{Sikorski1989} 
found that the energies of electronic states in InSb quantum dots effectively 
depend on the magnetic field strength, but their study was restricted to low 
lying energy states of very deep quantum dots. This is remarkable, since the 
first clear evidence of  discrete electronic states in semiconductor 
nanostructures was found a year earlier by Reed {\em et al.} \cite{Reed1988}.

There are numerous theoretical papers that deal with the subject of binding of resonances  and the physical traits that characterize the phenomenon in nanostructures, for instance Buczko 
and Bassani~\cite{Buczko1996} analyzed the bound and resonance states of spherical 
GaAs/$\mbox{Ga}_{1-x}\mbox{Al}_x \mbox{As}$ quantum dots, latter on
Bylicki and W. Jask\'olski \cite{Bylicki1999} analyzed the binding of one-electron
resonances in a  semiconductor quantum dot model. They found that the width 
of shape resonances were non-increasing functions of the magnetic
field strength, and that for large enough values the width becomes null.
Resonance states of two-electron systems, without magnetic fields, were
analyzed in a quantum dot~\cite{Bylicki2005} and atomic systems~\cite{Chakraborty2011}. 
Also, a two-electron quasi-one-dimensional system was studied using 
entanglement quantities~\cite{Kuros2015}. Sajeev and Moiseyev~\cite{Sajeev2008} 
demonstrated that the lifetime of resonance states of two-electron spherical 
quantum dots can be controlled by varying the confinement strength, Genkin 
and Lindroth~\cite{Genkin2010} reported that such control can be compromised by Coulomb 
impurities. Ramos and Osenda~\cite{Ramos2014} analyzed the
resonance states of one-electron cylindrical quantum dots with an external magnetic 
field using two quantum information-like quantities, the fidelity and the localization probability, {\em i.e.}, the 
probability that the electron is inside the potential well, to characterize 
the binding phenomenon. It is worth to mention again that  the studies mentioned in this paragraph employ the EMA. 

In this paper we use the $\mathbf{k}\cdot\mathbf{p}$ method to calculate the band structure of nanowires with a quantum dot embedded in them. We consider the transition between bound states and resonance states, which is  driven by applying a magnetic field along the nanowire axis. The single electron can be confined in an electrostatically defined quantum dot, or in a quantum dot made of materials different from that of the nanowire where it is embedded. In particular, we are interested in scenarios with different physical features than those found in nanowires whose phenomenology is adequately modeled with the EMA. We show how a resonance state can be detected using the {\em localization probability} \cite{Ramos2014,Garagiola2018}, and how the presence of a magnetic field affects degenerated states. This detection method has the advantage that does not depend on complex algebra calculations, as is the case in the complex rotation  \cite{Moiseyev1998} or the  complex absorbing potential \cite{Sajeev2009} methods.  

The paper is organized as follows, in Section~\ref{sec:model} we present the $\mathbf{k}\cdot\mathbf{p}$ Hamiltonian, the models for quantum dots embedded in semiconductor nanowires and the numerical methods used in this work to obtain the band structure that will be studied later on.  Many technical  details about the Hamiltonian and  the general expression of  the matrix elements  involved in the Rayleigh-Ritz method, are deferred to Appendixes~\ref{ap:Hamiltonian} and \ref{ap:matrix-elements}. Also in Section~\ref{sec:model}, we discuss the theoretical framework used to detect the presence of resonance states in numerically calculated spectra.  The results obtained for the model with a  quantum dot induced by an electrostatic confining potential are presented in Section~\ref{sec:Results}, while the results corresponding to the model with a semiconductor quantum dot  are presented in Appendix~\ref{ap:materials}. Finally, in Section~\ref{sec:conclusions} we discuss the results and present some future perspectives.

\section{\label{sec:model}Hamiltonian, models and methods}

\subsection{Hamiltonian}
To study the band structure of a semiconducting nanowire we  employ an 8-band {\bf k.p} Hamiltonian, which takes into account electron and hole bands.  Since the system may have an external magnetic field ($B$) and a confinement potential ($V$) applied, the total Hamiltonian that we will consider is given by
\begin{equation}\label{eq:Hamiltonian}
    H = H_{KL} + H^B+g\mu_B\kappa B + V
\end{equation}
where $H_{KL}$ is the 8-band Kohn-Luttinger Hamiltonian \cite{Kishore2014,tomic},  $H^B$ is the magnetic field term resulting from the Peierls substitution  \cite{Novik2005,Luttinger1955,Hofstadter1976} and $g\mu_B\kappa B$ is the Zeeman term. 
The potential $V$ takes into account the effect produced by  electrostatic potentials applied to the nanowire. In the case that the nanowire is homogeneous, $V$ could break the translational invariance of the system. 

We consider a zincblende nanowire whose axis coincides with the growth direction $\left[ 001\right]$. Many of the material parameters needed in the Hamiltonian, Eq. \ref{eq:Hamiltonian}, can be found in References \cite{Kishore2014,Kishore2012} and in Reference \cite{Saidi2010}, where the theoretical values calculated for many parameters are compared with the corresponding experimental values. It is worth to point that the Kane energy that we use is calculated in this last Reference. The authors found an excellent agreement between the theoretical and experimental values, note that we choose to use the experimentally obtained values for the Luttinger parameters see Table~\ref{tabla1} and Reference \cite{Saidi2010}.

The Hamiltonian in Equation \ref{eq:Hamiltonian} has been extensively used to study different kinds of nanowires. For instance, is the one used by Kishore and collaborators in \cite{Kishore2014} to analyze the behavior of the band structure of free standing nanowires. For a complete expression of the Hamiltonian and the operators and parameters involved in it see Appendix~\ref{ap:Hamiltonian}. 

We aim to study two different types of nanowire. The first one consists in a homogeneous one-material nanowire with a confining potential applied, which defines a potential well and barriers inside the nanowire. The second type consists on a nanowire with other semiconductors embedded in it.  In this case, the confining potential is equal to zero.

\subsection{Model one}
\begin{figure}[hbt]
\includegraphics[width=0.7\linewidth]{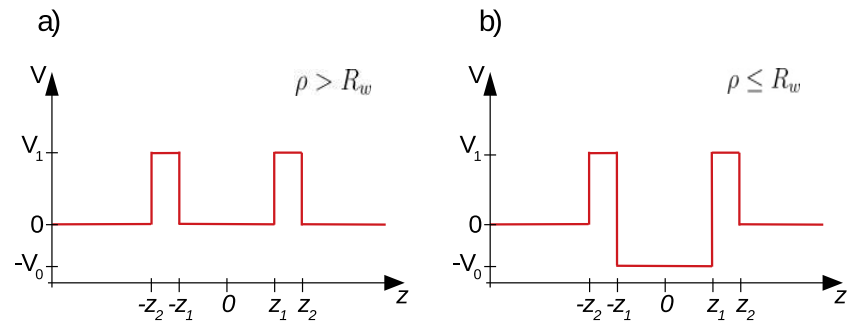}
\caption[Cartoon of the electrostatic confining potential]{The figure shows the confinement potential $V$,  as a function of the coordinate along the cylinder axis $z$,  for a) $\rho>R_w$  and b) $\rho\leq R_w$.
}\label{fig:perfiles-pot}
\end{figure}

The nanowire in this model is a cylinder of external radius $R$,  made of a single semiconductor material. We consider a cylindrical coordinate system, where the longitudinal coordinate, $z$, coincides with the nanowire axis. The confining potential is given by
\begin{eqnarray}
\label{eq:potential}
    V(\rho,z)= \begin{cases}
    -V_0\, &\mbox{ if } \, |z|\leq z_1\mbox{ and } \rho \leq R_w   \\
    V_1\, &\mbox{ if } \,z_1<|z|\leq z_2 \\
    0 &\mbox{ else }
     \end{cases}
\end{eqnarray}

this is, $V$ consists of a cylindrical potential well of radius $R_w<R$, length $2z_1$ and depth $-V_0$, and two potential barriers of radius $R$, length $z_2-z_1$ and height $V_1$,  at both sides of the potential well. The potential well and barriers define a quantum dot-like structure.

Figure~\ref{fig:perfiles-pot} shows the potential profile as a function of the longitudinal coordinate $z$, for a) $\rho>R_w$ and  b) $\rho\leq R_w$.

\subsection{Model two}

\begin{figure}[hbt]
\includegraphics[width=0.7\linewidth]{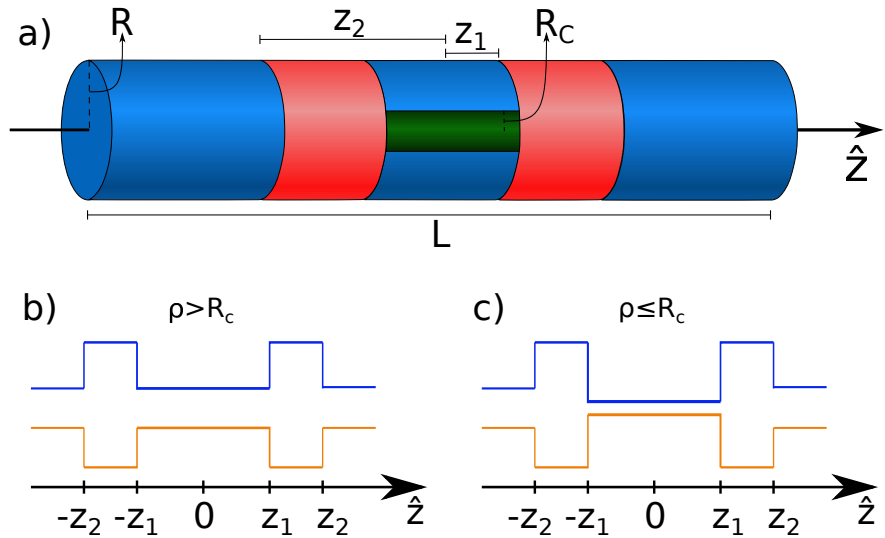}
\caption[Cartoon of the semiconductor quantum dot]{a) Cartoon of the second model considered. The nanowire is depicted with the blue cylinder of radius $R$ and total length $L>>z_1,z_2$. Embedded on it, there are a core cylinder of radius $R_c$ and length $2z_1$, shown as a green cylinder. Also, there are two embedded red cylinders of radius $R$ and length $z_2-z_1$. Each color represents a different material, that are chosen to define a quantum dot which is generated by the difference in the band offsets of the materials. Panels b) and c) show the conduction band bottoms (blue) and valence bands tops (orange) as a function of the coordinate along the cylinder axis $z$, for b) $\rho>R_c$ and c) $\rho\leq R_c$.
}\label{fig:cartoon2}
\end{figure}

The second system consists of a nanowire with three semiconductor cylinders embedded in it, and with $V=0$. These three cylinders are  made of two different materials, in a configuration that mimics the potential profile  of Equation~\ref{eq:potential}, by means of the band offsets between the materials involved. A cartoon of the nanowire is shown in Figure~\ref{fig:cartoon2} a), where each color represents a different material. The green cylinder has radius $R_c$ and length $2z_1$, while the red cylinders have radius $R$ and length $z_2-z_1$. The semiconductor materials of the red and green cylinders are chosen to form a well and barrier structure analogous to the potential in Equation~\ref{eq:potential}. Figure~\ref{fig:cartoon2} also shows the conduction band bottoms and valence bands tops profiles for b) $\rho>R_c$ and c) $\rho\leq R_c$. The details about the materials parameters necessary to implement the $\mathbf{k}\cdot\mathbf{p}$ method can be found in Appendix~\ref{ap:Hamiltonian}.

Comparing the potential profiles in Figure~\ref{fig:perfiles-pot} and the conduction band profiles in Figure~\ref{fig:cartoon2}, it is clear that an electron in the conduction band experiments a similar confining in both systems. The first model is simpler to treat since there are not changes in the material parameters, changes that are present in the second model.

The results for the two systems described above were obtained using the same approach. In the first model we studied the band structure near the bottom of the conduction band by changing the  parameters of the potential: the length and the depth of the potential well, and the length and the height of the potential barriers. In the second model, the band offsets are fixed for a given set of materials, so the only parameters that can be changed are the lengths $z_1$ and $z_2$, and the radius $R_c$. We used these parameters to regulate the number of states in the gap below the bottom of the conduction band. In both cases we looked for scenarios with well isolated resonance states that lied near the bottom of the conduction band.

\subsection{Rayleigh-Ritz variational method}

In both models we applied the Rayleigh-Ritz variational method to obtain the spectrum and eigenstates of the systems. This method has been used in a range of situations that go from band structure of free standing nanowires \cite{Kishore2014,Kishore2012} and resonance states \cite{Ramos2014,Garagiola2018,Ferron2009,Pont2010} to band structure properties of quantum wells \cite{Krishtopenko2018}. For a given Hamiltonian $\mathcal{H}$, and a test function $\phi^t$, which can be written as 
\begin{equation}
\phi^t = \sum_{\nu} c_{\nu} \varphi_{\nu} ,
\end{equation}
where $\lbrace \varphi_{\nu} \rbrace_{\nu=1}^Q$ is a basis set with  $Q$ elements. The Rayleigh-Ritz method minimizes the normalized expectation value of the Hamiltonian
\begin{equation}
\tilde{E} = \frac{\left(\phi^t, \mathcal{H} \phi^t\right)}{(\phi^t,\phi^t)},
\end{equation}
with respect to the coefficients $c_{\nu}$, this minimization procedure results in an algebraic eigenvalue problem. Some times, the basis set functions may depend on a set of parameters, which can take different values and are commonly termed as "non-linear variational parameters" \cite{Nesbet}.

For the models that we consider, the basis set used takes into account the dependence on the radial, angular, and longitudinal along the axis of the nanowire coordinates, and the boundary conditions. Solving the variational eigenproblem
\begin{equation}
\label{eq:eigenproblem}
\mathrm{H} \psi^v = E^v \psi^v ,
\end{equation}
where $\mathrm{H}$ is the matrix representation of the Hamiltonian in
Equation~\ref{eq:Hamiltonian}, in some function basis set $\psi^v$, 
provides a discrete spectrum which is a good approximation for the energy sub bands around the semiconductor gap, and for discrete states inside the gap. The states in the different energy bands are extended states, as those in  bulk material, while those inside the gap are well localized states,  with probabilities negligibly small  of finding the particle far from the potential well or the material well.  We will use this property to distinguish between extended and localized approximate states, since the application of the Rayleigh-Ritz method results in a discrete spectrum of variable density, in which it is not always evident whether an eigenvalue corresponds to one or another type of state.

The test functions employed in the Rayleigh-Ritz method can be written as a linear combination of the basis functions
\begin{equation}
\label{eq:basis}
 \psi_{n\,m\,l}(\rho,\varphi,z)=
 \frac{e^{i m\phi}}{\sqrt{2\pi}} \, \frac{\sqrt{2}}{RJ_{m+1}(\alpha_{ml})}J_m\left(\alpha_{ml} \frac{\rho}{R} \right) \,\, 
\sqrt{\frac2L}\cos\left(n \pi \frac{2z}{L}\right)  ,
\end{equation}
where $J_m$ are the Bessel functions of the first kind, $\alpha_{ml} $ is the $l$th zero of the function $J_m(\rho)$, $m=0,\pm1,\pm2,...,\pm M$, $l=1,2,...,N_z$ and $n=1,2,...,N$.

To obtain the band structure we considered different sets of values for $M,N_z$ and $N$. First we re-calculated the results obtained in \cite{Kishore2014}, where the authors use $N_z=10$ and $M=4$ to study the energy spectrum of a core-shell nanowire, using the 8-band Kohn-Luttinger Hamiltonian that we use in this work. The authors of Reference~\cite{Kishore2014} assure that this values for $N_z$ and $M$ are sufficient for convergence of the energy levels. We repeated the calculation  to test our algorithms, and saw that the states near the bottom of the conduction band can be accurately obtained  using only  zero angular momentum basis states, {\em i.e.}, $M=0$. Since the states near the conduction band bottom  are the ones that we want to study, we adopt for further calculations $N_z=10$ and $M=0$. To study systems without transational invariance in $z$, a basis set functions depending on the $z$ coordinate has to be included. Calculating the spectrum of the systems as a function of the basis size $N$, we found that for $N\geq80$ the change on the energy levels when increasing $N$ is smaller than $10^{-5}$.


In Appendix~\ref{ap:matrix-elements} we present some helpful expressions and intermediate results to obtain the matrix representation of the Hamiltonian in order to solve Equation \ref{eq:eigenproblem}.

\subsection*{Removing degeneracies with magnetic fields}

Before presenting the results obtained using the $\mathbf{k}\cdot\mathbf{p}$ method, it is convenient to remember how the application of a magnetic field affects a band of energy levels that are doubly degenerate because they have been obtained from a spin-independent Hamiltonian, as is the case in the simplest EMA approximation \cite{Ramos2014,Garagiola2018,Bylicki1999,Bylicki2005}, or from a spin-dependent Hamiltonian without external magnetic fields \cite{Konig2007, Zhou2008,Durnev2016,Raichev2012}. 

\begin{figure}[hbt]
\includegraphics[width=0.4\linewidth]{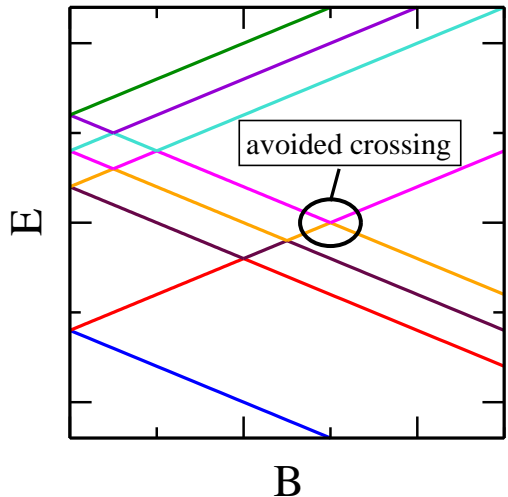}
\caption[Cartoon of the behavior of degenerate states versus B]{Schematic representation of the behavior of degenerate energy levels when an external magnetic field is applied. For $B=0$ there are four doubly-degenerate energy levels. When the magnetic field strength is not zero the degeneration is removed. The  ellipse is used to signal one of the multiple avoided crossings that appear in the spectrum if the energy  levels are close enough.
}\label{fig:cartoonB}
\end{figure}

The application of a magnetic field to a given system adds  to the Hamiltonian the Zeeman term, that is proportional to the strength of the field, and terms that come from the canonical momentum which now includes the vector potential $\vec{A}$. This last term is included in the $\mathbf{k}\cdot\mathbf{p}$ Hamiltonian (and other model Hamiltonians) through the Peierles substitution.

When the field intensity is small, the Zeeman term dominates the behavior of the energy levels, and each doubly degenerate energy level at null magnetic field unfolds into two levels that vary linearly with the field strength, $B$. The separation between the now non-degenerate levels grows and, eventually, these two levels can become nearly degenerate with others, producing a series of avoided crossings as shown in the cartoon in Figure~\ref{fig:cartoonB}. This cartoon is an schematic representation but it is qualitatively correct, as it will be shown by the Figures included in the next Section.   For larger field strengths all the other terms depending on the magnetic field become more and more important, giving place to the non-linear behavior of the energy levels. 

\subsection*{ Identifying resonance state energies with localization probabilities}
In this subsection we aim to discuss  the qualitative behavior of the energy levels of a system with a localized bound state and a band of extended states as, for instance, a few-electron quantum dot near the ionization threshold, {\em i.e.} the scenario when the bound state enters into the band. We are considering energy levels that are obtained using an approximate method like the Rayleigh-Ritz variational one.   In these quantum dots, changing the value of a parameter of the Hamiltonian, let us say the depth of the potential well where the bound states are localized, $\alpha$, leads to the loss of one electron, {\em i.e.} the {\em ionization} of the system. The Rayleigh-Ritz method renders the continuum of extended states into a bundle of energy levels that do not depend on this parameter {\bf except} when their energies are close to the energy of the resonance state, that is the continuation of the bound state embedded in the continuum \cite{Ramos2014,Ferron2009,Pont2010}.

\begin{figure}[hbt]
\includegraphics[width=0.4\linewidth]{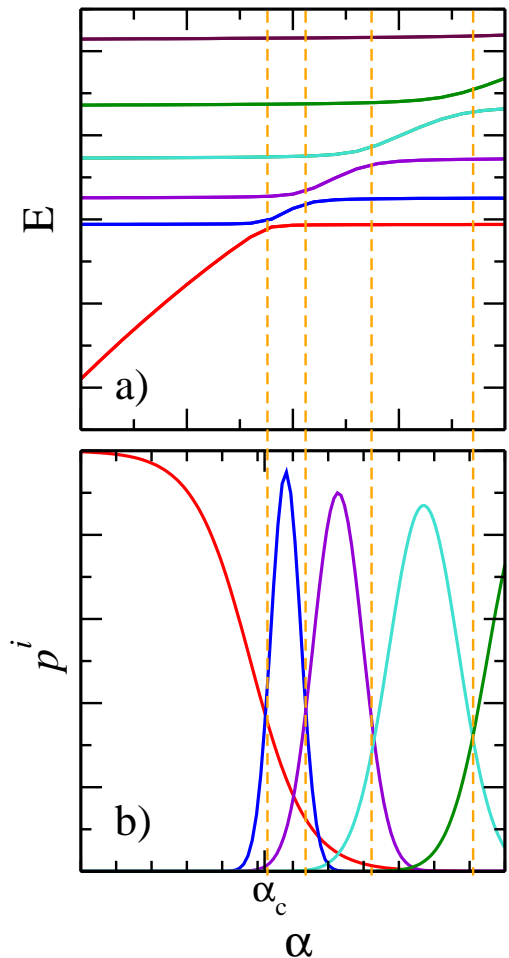}
\caption[Cartoon of the energy spectrum and localization probability near the ionization threshold]{a) Schematic spectrum of a system near the localization threshold {\em vs } the parameter $\alpha$. The solid red line corresponds to a bound state that enters into a band, which is shown using five energy levels for $\alpha<\alpha_c$, and with six energy levels for $\alpha>\alpha_c$. $\alpha_c$ is the value for which the ionization of the system occurs.  b) The localization probability, $p^i(\alpha)$ versus the same parameter. The color code used show which probability corresponds to which energy eigenstates.  The orange dashed lines are included as a  guide to the eye to show where the avoided crossings $\tilde{\alpha}_{i,i+1}$ are located. It is clear then that the states in the band are highly localized only when its energy is close to the resonance energy, {\em i.e.} between two successive avoided crossings.
}\label{fig:cartoonE}
\end{figure}

The scenario described above is schematically depicted by the cartoon in figure~\ref{fig:cartoonE} a), which shows the typical behavior of the energy levels versus $\alpha$ in the neighborhood of a {\em ionization threshold}, that occurs for $\alpha=\alpha_c$. The cartoon depicts the situation where a state, whose energy is shown with a solid red curve, goes from bound state for $\alpha<\alpha_c$ to an extended state for $\alpha>\alpha_c$. The other eigenvalues are repelled by the state that enters into the band, and a number of avoided crossings appear in the spectrum. The resonance dependence with $\alpha$ is signaled by these successive avoided crossings. So, calling $\tilde{\alpha}_{i,i+1}$ the  value of $\alpha$ where the avoided crossing between energy levels $E^i(\alpha)$ and $E^{i+1}(\alpha)$ is located, it has been shown that the resonance energy $E_{res}(\alpha)$ is,  with a very high accuracy, equal to the energy of a variational eigenvalue, $E_{res}(\alpha) \sim E^i(\alpha) $, for those values of  $\alpha \in (\tilde{\alpha}_{i-1,i}, \tilde{\alpha}_{i,i+1} )$ \cite{Ramos2014,Garagiola2018,Pont2010}.

A more complete characterization of resonance states can be obtained using not only the spectral information, but studying the information stored in the eigenfunctions. In this sense, the variational eigenfunctions  can  be used to track down the resonance state by calculating the probability that they are localized in the spatial region where bound states can be expected, {\em i.e.} in  the potential well of a quantum dot model. This probability can be calculated as
\begin{equation}\label{eq:prob-loc-tipica}
p ^i(\alpha) = \int_{V_L} |\phi^i_{\alpha} (\mathbf{x})|^2\, d\mathbf{x} .
\end{equation}
where $V_L$ is the region where the potential well is located, $\phi^i_{\alpha}$  is a variational eigenfunction calculated for a given value of the parameter $\alpha$, with an energy $E^i(\alpha)$ that is close to the resonance energy. The resonance state can be singled-out plotting $p^i(\alpha)$ and the spectrum $E^i_{\alpha}$, since the resonance state is well localized inside the well. In the next paragraph we present a qualitative example of the detection procedure.

 In Figure~\ref{fig:cartoonE} b) it is shown the qualitatively behavior  of $p^i(\alpha)$ for the eigenfunctions corresponding to the eigenvalues shown in a).  The  state whose energy and probability is shown using a red solid line, has a large probability of being localized for $\alpha<\alpha_c$, but when it enters into the continuum it becomes extended and, consequently, its probability drops abruptly to almost zero. The states in the band have a very small probability of being localized, except near the resonance state. The color code used in both panels indicates which eigenvalue and eigenfunction correspond to each probability.

\section{\label{sec:Results}Results}
In the following we will focus on the spectral properties of systems with potential well and barriers generated through the application of electrostatic potentials, and differ the presentation of results obtained for systems with well and barriers constructed with different semiconductor materials to Appendix~\ref{ap:materials}.

\subsection{\label{subsec:B.eq.0}Spectra for model one type systems with $B$=0}

In this sub-section and in the following we will focus in the approximate spectrum that is obtained solving the numerical eigenvalues problem. In particular, we look for values of the potential well depth and barriers height energies that result in the appearance of localized states below the conduction band. As our results show, the number of these localized states can be easily controlled by choosing appropriate values for these two parameters.

\begin{figure}[hbt]
\begin{center}
\includegraphics[width=0.45\linewidth]{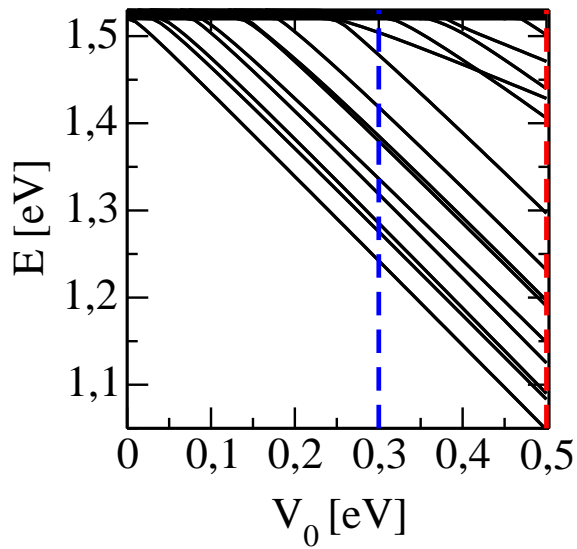}
\end{center}
\caption[Energy spectrum versus the depth of the well for the electrostatic quantum dot]{Energy spectrum as a function of the depth of the well, $V_0$, for a GaAs nanowire of radius $R=15$~nm, and potential parameters $z_1=3.5$~nm, $z_2=4.75$~nm, $R_w=5$~nm and $V_1=0$~eV. The dashed vertical lines show the values of $V_0$ chosen for further study of the system. The blue dashed line corresponds to $V_0=0.3$~eV, and the red dashed one to $V_0=0.5$~eV.
}\label{fig:pot-V0}
\end{figure}

\begin{figure}[hbt]
\includegraphics[width=0.8\linewidth]{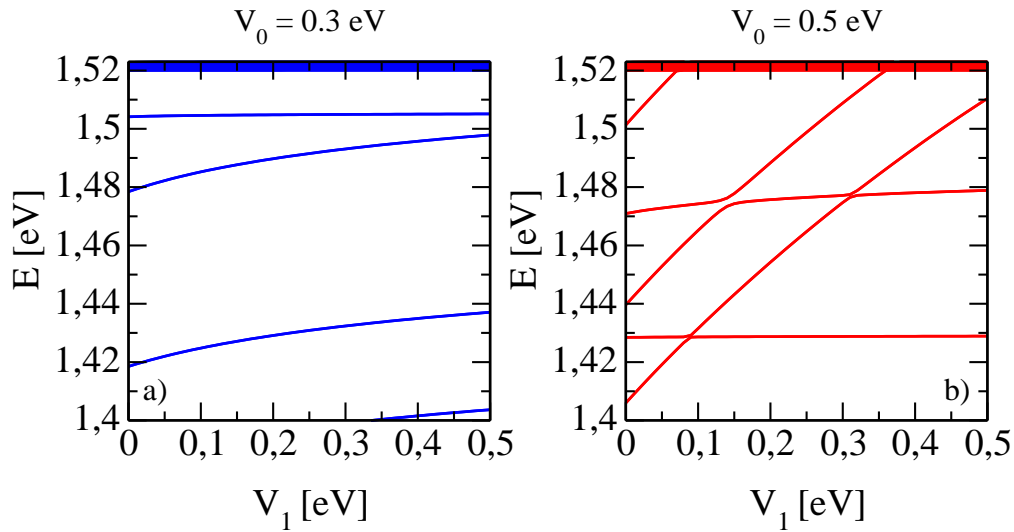}
\caption[Energy spectrum versus the height of the barriers for the electrostatic quantum dot]{Energy spectrum as a  function of the height of the barriers, $V_1$, for a GaAs nanowire of radius $R=15$~nm, and potential parameters $z_1=3.5$~nm, $z_2=4.75$~nm and $R_w=5$~nm. The left panel, in blue, corresponds to $V_0=0.3$~eV, and the one in the right, in red, to $V_0=0.5$~eV.
}\label{fig:pot-V1}
\end{figure}

Figures~\ref{fig:pot-V0} and \ref{fig:pot-V1} show the discrete spectrum that appears near the bottom of  the conduction band when a potential well is electrostatically induced in a GaAs nanowire. Note that the bottom of the conduction band is located around $1.52$ eV, and that for energies above that the energy density of levels is larger. In Figure~\ref{fig:pot-V0} the energy of the eigenvalues is plotted against the  potential well depth $V_0$, for $V_1=0$. The radius of the nanowire is $R=15$~nm. The vertical dashed lines mark particular values of the potential well depth, that will be further analyzed in Figure~\ref{fig:pot-V1}, as shown by the color used in each panel to depict the eigenvalues. In Figure~\ref{fig:pot-V1} the eigenvalues are shown as functions of the potential barrier height $V_1$. 

As can be appreciated in the right panel of Figure~\ref{fig:pot-V1}, there is a number of eigenvalues that belong to the discrete spectrum when $V_1=0$ that enter into the conduction band  (CB) for some value of $V_1$. These eigenvalues are of particular interest since they are the origin of resonance states once they enter into the CB, so the neighborhood of these particular values of $V_0$ and $V_1$  is an appropriate place to look for binding of resonances when an external magnetic field is applied.

It is interesting to note that the eigenvalues of the discrete vary almost linearly as a function of $V_1$, but there are groups with very different slopes. This is due to the different symmetries of the corresponding states, which makes some of them more susceptible to the change in the height of the barrier of potential than the others. On the other hand, the proximity between the eigenvalues below the conduction band bottom (CBB) and its, relatively, large number, could obscure the analysis of the ensuing physical phenomena. Fortunately, the number of discrete eigenvalues below the CBB can be further reduced by choosing smaller radius for the well, while keeping the values of $V_0$ and $V_1$ in the same range than those used to obtain the data in Figures~\ref{fig:pot-V0} and \ref{fig:pot-V1}. 

We have presented results only for a confining potential with $R_w=5\mbox{nm}$, and $z_1=3.5$~nm, and $z_2=4.75$~nm, but we have also performed calculations for many other sets of these lengths. In particular we analyzed different radius of the potential well, such as  $R_w=7\mbox{nm}$ and  $R_w=9\mbox{nm}$. For  these other cases the binding of resonances is also present. The values of  the parameters $V_0$, $V_1$, are similar to those used for the $R_w=5\mbox{nm}$ case.

\subsection{\label{subsec:B.neq.0}Spectra for model one type systems with $B\neq0$}

Before proceeding with the presentation of the results about the spectrum properties, it is worth to point out that the presence of resonance states can be detected using a number of methods such as complex rotation, complex absorbing potentials, density of states among others, see References~\cite{Ramos2014, Bylicki1999,Bylicki2005,Sajeev2008,Sajeev2009,Pont2010}. All these methods have a long history, and have been used to successfully calculate the energy and width of resonance states when the spectrum properties are calculated using  Schrödinger-like equations. 

There are not, to our knowledge, implementations of these methods to systems whose description is based on the $\mathbf{k}\cdot \mathbf{p}$ method, so to detect the presence of resonance states we will rather use the previously introduced localization probability \cite{Ramos2014}, now called $P_{w}$, which is given by
\begin{equation}
P_w(\psi^v) = \int_{well}   |\psi^v(\mathbf{x})|^2 \, d\mathbf{x} ,
\end{equation}
where $|\psi^v(\mathbf{x})|^2$ is the square modulus of the eight-component spinor, that is an eigenvector of Equation~\ref{eq:eigenproblem}. As it will be shown, for the results obtained with the $\mathbf{k}\cdot\mathbf{p}$ method $P_{w}$ acts very much like an order parameter. It is able to detect when a localized state appears below the bottom of a continuum,  therefore signaling the binding of a resonance when an increasing  magnetic field is applied to a system with a resonance state. 

\begin{figure}[hbt]
\includegraphics[width=1\linewidth]{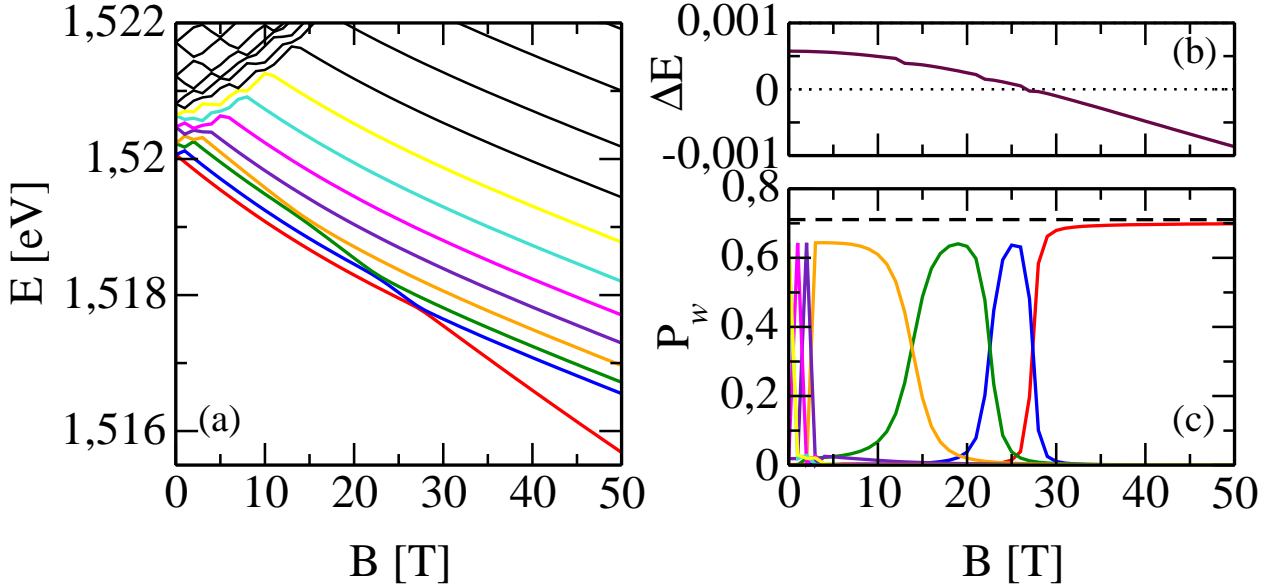}
\caption[Energy spectrum, $\Delta E$ and localization probability versus $B$ for the electrostatic quantum dot]{a) Energy spectrum as a function of the external magnetic field, $B$, for a GaAs nanowire of radius $R=15$~nm and potential parameters $z_1=3.5$~nm, $z_2=4.75$~nm, $R_w=5$~nm, $V_0=0.5$~eV and $V_1=0.3625$~eV. b) Difference in energy, $\Delta E$, between the highest localized state near the CBB and the CBB state, as a function of $B$. c) The localization probability versus $B$ for some of the states in panel a). The color code used in both panels is the same, in order to facilitate the identification of the localization probabilities with the corresponding states. The dashed black line shows the localization probability for the higher energy state in the gap.
}\label{fig:pot-B}
\end{figure}

Figure~\ref{fig:pot-B} a) shows the behavior of the variational spectrum near the bottom of the conduction band as a function of the field strength $B$, for potential parameters $V_0=0.5$~eV, $V_1=0.3625$~eV, $z_1=3.5$~nm, $z_2=4.75$~nm and $R_w=5$~nm.  From Figure~\ref{fig:pot-V1}, it is clear that a state in the gap enters into the conduction band for $V_1\sim 0.36$~eV, so for larger values of $V_1$ there must be a resonance state lying very close to the bottom of the conduction band. All the eigenvalues are doubly degenerate for $B=0$, and this degeneracy  is broken when the magnetic field is applied, as described above. Two curves emerge from each of the doubly degenerate eigenvalues at $B = 0$, one increasing and the other decreasing as functions of the field intensity. 

In panel a) it can be seen that there is a group of "parallel" curves, which correspond to decreasing curves, that do not have intersections avoided with each other, while, near the vertical axis, there is a series of curves with numerous intersections avoided, which occur between increasing  and decreasing eigenvalues. In any case, there is an almost straight decreasing curve that shows a different behavior from the others, that corresponds to the resonance state that becomes a localized bound state for $B\sim 25T$. Panels b) and c) have the accessory information that allows us to single out that this eigenvalue is the one that corresponds to the resonance state-bound state transition.

Studying the localization probability for all the states near the bottom of the conduction band, it is possible to identify the states that are extended from those that are localized, since the former have close to zero localization probability and the later have localization probability values that are close to the values shown by in-gap states that lie far from the CBB, where far from the CBB means energy differences larger than $0.01$~eV. 

The difference between a resonance state and an isolated bound state, from the point of view of the localization probability (LP), is given by the existence, or not, of an extended state below it. Panel b) of Figure~\ref{fig:pot-B} shows the difference in energy between the state near the CBB that possesses a high LP, $E_{HLP}$, and the lowest lying extended state, $E_{CBB}$, so $\Delta E$ can be written as
\begin{equation}
    \Delta E = E_{HLP}(B) - E_{CBB}(B)= \left\lbrace
    \begin{array}{lcl}
    E_{res}(B) - E_{CBB}(B), & \mbox{if the HLP state is in the band}\\
    & \\
    E_{loc}(B) - E_{CBB}(B), & \mbox{if the HLP state is in the gap}\
    \end{array}
    \right..
\end{equation}

There are other bound states in the gap, but they can be found for energy differences, with respect to the CBB,  larger than $0.01 \mbox{eV} \gg \Delta E$, {\em i.e.} when the resonance state under study becomes a bound state it is well isolated from other in-gap bound states. Also, it is well isolated from other resonance states that can be present in the system when the transition occurs.  The transition from resonance ($\Delta E >0$) to bound state ($\Delta E <0$) can be clearly appreciated in Figure~\ref{fig:pot-B} b).

\section{Discussion and conclusion}
\label{sec:conclusions}

Our results show that the binding of resonances can be effectively implemented in semiconductor nanostructures embedded in nanowires in both cases considered, electrostatic quantum dots and material quantum dots. The effect seems similar to the one found using the EMA, at least for the materials that we considered in this paper. It would be interesting to consider combinations of other materials with  larger offset parameters, as this would result in a higher probability of localization and,  presumably, a stronger localization of the resonance. Also, it would allow to use larger lengths in the $z$ direction for the core cylinder and barriers, and a larger radius of the core, which would make the  lengths and radii values easier to implement  experimentally. 

The combination of materials that we have considered had multiple purposes. First, to allow us to compare how precise are our numerical results when compared with previous results found for similar systems \cite{Kishore2014,Kishore2012}, second to consider a combination of materials that are known to form wires and other nanostructures, and third to compare with previous results calculated using very much the same materials, but using the EMA \cite{Garagiola2018}.
Before calculating the spectrum of nanowires with potential well and barriers, we calculated the spectrum for core-shell nanowires as those considered in Reference~\cite{Kishore2014} and found and excellent agreement between both sets of results. Besides, we found that for only one material nanowires the energy of states near the bottom of the conduction band can be obtained with a very good approximation considering only basis set functions with angular momentum  component $L_z=0$. 

The details of the electrostatic gates that are used for generate the confinement potential do not seem to be important, as long as the localization in a well with an effective radius smaller than the external radius of the wire is feasible. Resonance binding has been found for many different types of binding potentials, and for one electron systems the only necessary condition is the presence of a potential well and barriers. So, there is no need to consider a piecewise constant potential but for analytical simplicity, moreover, it is the simplest potential with a profile similar to the conduction band profile considered in what we called model two. Assuming a potential well with cylindrical symmetry is necessary to simplify the basis set required to implement the Rayleigh-Ritz method. In the same sense, the precise longitudinal dimensions values of the barriers and wells were chose to single out and separate the resonance states present in the system. Our results, in both kind of models, show that the binding phenomenon can be found for an ample sets of materials and dimensions.

The Rayleigh-Ritz method has been successfully applied to a number of different scenarios, not only to the calculation of the energy of bound states. For instance, it has been  applied to study the band structure of model semiconductor nanostructures which are described using a $\mathbf{k} \cdot \mathbf{p}$ Hamiltonian. It has been found that the results describe, correctly, the physics near the bottom of the conduction band and the top of the valence band. This approach has been exploited in References \cite{Kishore2014} and \cite{Kishore2012} in the context of free-standing nano-wires, in References \cite{Garagiola2018} and \cite{Moiseyev1998} in the context of the calculation of mean life times of shape resonances, and in Reference \cite{Durnev2016} to obtain ``realistic'' two-dimensional helical states in quantum wells. Since the number of localized states inside the gap in the situations considered in this paper is fairly low the results of the variational method are accurate, as is the value of the gap energy and the values of the lowest  energy of the conduction  band. These elements are the only ones needed to identify the transition from bound to resonance state and back.

Determining the width of the resonance as a function of the field, when using the $\mathbf{k}\cdot\mathbf{p}$ method, is even more complicated than in the case of using the EMA. To begin with, we have to consider eight coupled equations, in the former case, instead of a single one  as in the later one. The size of the Hamiltonian matrix involved is on the order of several tens of thousands, with the consequent loss of numerical precision when the eigenvalues and eigenstates are calculated. Nevertheless, introducing a {\em soft-wall potential}, located at variable distances from the quantum dot, and using this distance as a variational parameter to obtain a density of states, we have estimated a resonance width of $\Gamma \sim 10^{-6} \mbox{eV}$. To determine the value of the width of a resonance,  it is commonly assumed that the density of states has a Lorentzian shape near the peak associated to the energy of the resonance. So far, our results do not show the required Lorentzian shape to proceed with the necessary fitting to determine $\Gamma$.

We choose to defer most  of the technical details to the Appendices to enhance the readability of our paper, and do not delay the presentation of results. Nevertheless, we want to emphasize that the analytical calculations presented in Appendix~\ref{ap:matrix-elements} could help other researchers to implement their own matrix elements calculation in a easier and faster way than ourselves. The expressions that we included pay a lot of care dealing with the matching conditions at the interfaces between materials and the proper conversion of the operators involved in the calculus to their coordinate representation. Because we considered a quantum dot embedded on a nanowire, the matching conditions at the interfaces of the materials must be taken into account in the $\rho$ and $z$ coordinates, which differs from previous results found in the literature \cite{Kishore2014,Kishore2012}. 

The binding of resonance states properties observed for  systems modeled 
using the $\mathbf{k}\cdot\mathbf{p}$ method are similar to the ones found 
using the Effective Mass Approximation, but the inclusion of the spin adds 
novel scenarios where some states make the transition from bound to resonance 
states, or vice versa, depending on its spin. From our results it is clear that, if for $B=0$ a degenerate bound state lies close enough to the CBB, then for $B\neq 0$  the state whose energy grows with the field strength will collide with the CBB at some value of $B$, then becoming a resonance. On the other hand, the other state which has a decreasing energy as a function of the field strength, will increase its distance with the CBB. This scenario, and the resonance to bound state that we discussed extensively, could provide control techniques for electrons in quantum dots.

\section*{Acknowledgements}

Both authors want to acknowledge partial financial support from Secyt-UNC, National University of Córdoba,  and CONICET, Argentina.


\appendix
\section{\label{ap:materials} Binding of resonances in a model two type systems}

Here we present the results for the model in which the quantum dot is obtained by a certain configuration of different materials (see Figure~\ref{fig:cartoon2}). The materials used for the system are $\mbox{Ga}_{0.51}\mbox{In}_{0.49}$P for the nanowire, GaAs for the core and $(\mbox{Al}_{0.5}\mbox{Ga}_{0.5})_{0.51}\mbox{In}_{0.49}\mbox{P}$ for the barriers. The parameters of the system are $R=15$~nm, $R_c=5$~nm and $L=200$~nm. Since the depth of the well and the height of the barriers are fixed and given by the band energies of the materials involved, the parameters we vary in order to separate and control the number of localized states are the width of the core and of the barriers.
\begin{figure}[hbt]
\includegraphics[width=0.5\linewidth]{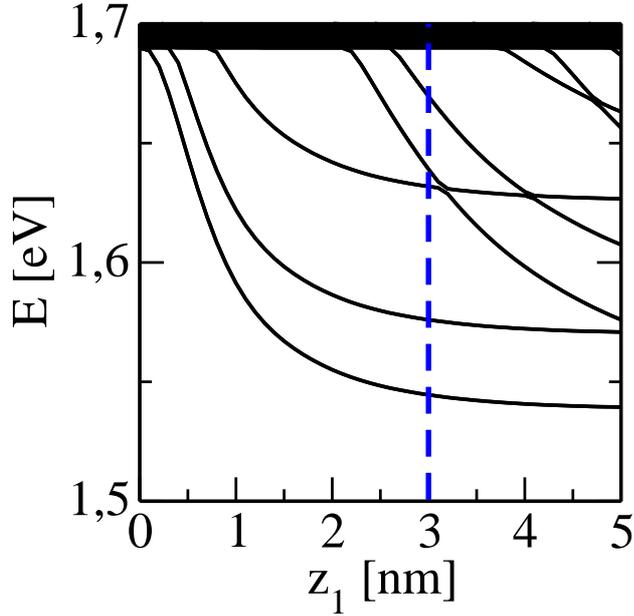}
\caption[Energy spectrum versus the core width, for the semiconductor quantum dot]{Energy spectrum as a function of the width of the core cylinder, $z_1$, for a $\mbox{Ga}_{0.51}\mbox{In}_{0.49}$P nanowire of radius $R=15$~nm, and GaAs core of radius $R_c=5$~nm. There are no barriers. The blue dashed line indicates the value $z_1=3$~nm, chosen for the next steps.
}
\label{fig:mat-z1}
\end{figure}
\begin{figure}[hbt]
\includegraphics[width=0.8\linewidth]{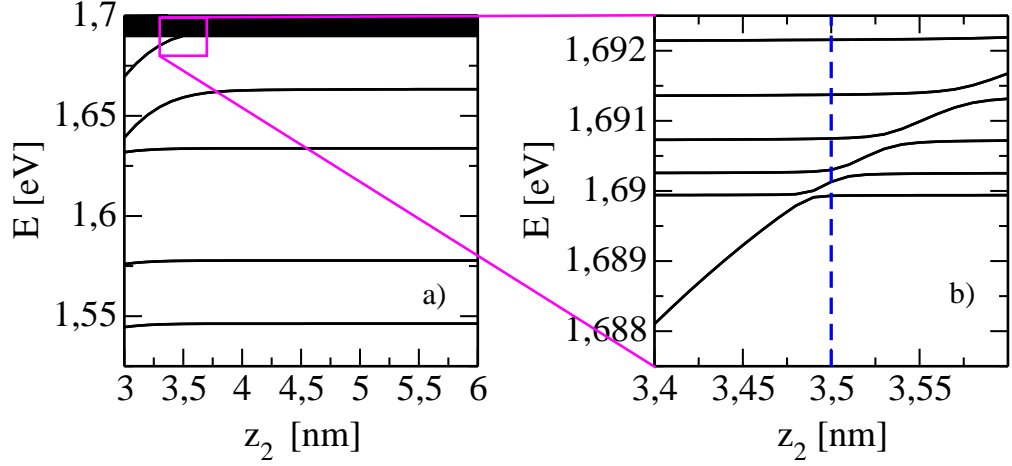}
\caption[Energy spectrum versus the barriers width, for the semiconductor quantum dot]{a) Energy spectrum as a function of $z_2$, for a $\mbox{Ga}_{0.51}\mbox{In}_{0.49}$P nanowire of radius $R=15$~nm, GaAs core of radius $R_c=5$~nm and width $z_1=3$~nm, and $(\mbox{Al}_{0.5}\mbox{Ga}_{0.5})_{0.51}\mbox{In}_{0.49}\mbox{P}$ barriers. b) Zoom of the a) panel near the ionization threshold. The blue dashed line indicates the value $z_2=3.5$~nm.
}\label{fig:mat-z2}
\end{figure}

Figure~\ref{fig:mat-z1} shows the energy levels as a function of $z_1$, for a system without barriers. In this case the dependence of the energy levels in the gap is not linear but, as in the case of the electrostatically produced quantum dot, two different behaviors can be detected. The blue vertical line marks the value of $z_1$ for which we add barriers to the system. In Figure~\ref{fig:mat-z2} the spectrum of the system is plotted against $z_2$, for $z_1=3$~nm. The right panel is a zoom of the area contained by the pink box on the left panel, and shows how the localized state of higher energy enters into the conduction band for $z_2\sim3.47$~nm. 

\begin{figure}[hbt]
\includegraphics[width=0.8\linewidth]{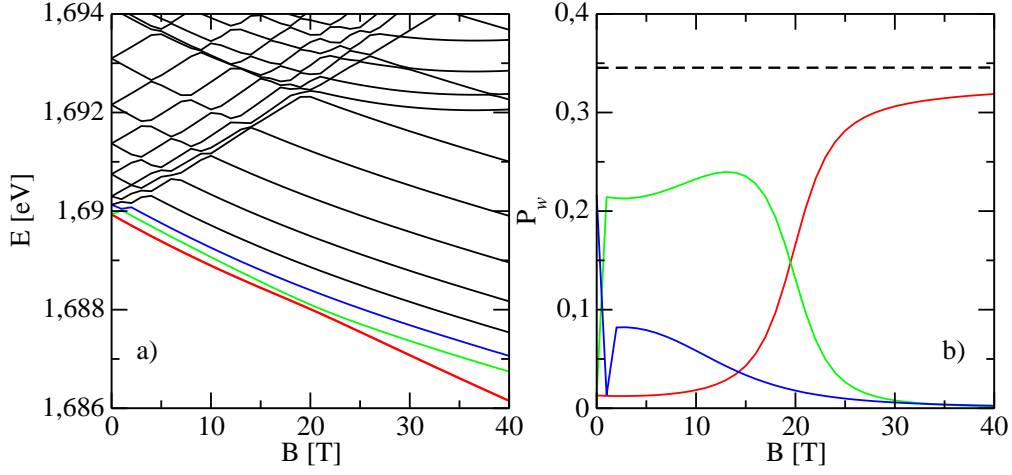}
\caption[Energy spectrum and localization probability versus the magnetic field strength, for the semiconductor quantum dot]{a) Energy spectrum as a function of the strength of the external magnetic field, $B$, for a $\mbox{Ga}_{0.51}\mbox{In}_{0.49}$P nanowire of radius $R=15$~nm. The embedded quantum dot is formed with a GaAs core of radius $R_c=5$~nm and  $(\mbox{Al}_{0.5}\mbox{Ga}_{0.5})_{0.51}\mbox{In}_{0.49}\mbox{P}$ barriers. The length parameters for the core and barriers are $z_1=3$~nm and $z_2=3.5$~nm. b) The localization probability versus $B$ for some of the states showed in panel a). The color code used in both panels is the same, in order to facilitate the identification of the localization probabilities with the corresponding states. The black dashed line in panel b) corresponds to the localization probability for the higher energy state in the gap. For $B=0$ the energy of this state is approximately 1.66 eV.
}\label{fig:mat-B}
\end{figure}

Let us consider now the effect of applying an external magnetic field for the system with $z_1=3$~nm and $z_2=3.5$~nm. Figure~\ref{fig:mat-B} shows that the behavior of the eigenvalues is qualitatively the same as the one described in Section~\ref{subsec:B.neq.0}, with respect to Figure~\ref{fig:pot-B}. In the left panel, which presents the spectrum as a function of the magnetic field, a state that depends almost linearly with the field strength can be seen. This resonance state exits the conduction band for $B\sim20$~T. The right panel shows the localization probability of this state as a function of $B$. The LP increases with the field strength, and is smaller than the LP of the highest energy state in the gap, depicted by a black dashed line.

\section{\label{ap:Hamiltonian} The eight-band Hamiltonian}
As stated in Section~\ref{sec:model}, the Hamiltonian used for the study of both models can be written as
\begin{equation}
        H = H_{KL} + H^B+g\mu_B\kappa B + V.
\end{equation}
The {\bf k.p} Hamiltonian includes information about the band structure and symmetry of the material, and  it is customary to write it down in the angular momentum basis $|J,m_z\rangle$:
\begin{eqnarray}
\label{eq:kpbasis}
|CB_+\rangle &=& |\frac12 , +\frac12\rangle = |s ;\uparrow \rangle \nonumber \\
|CB_-\rangle &=& |\frac12 , -\frac12\rangle = -|s ,\downarrow \rangle \nonumber \\
|HH_+\rangle &=& |\frac32 , +\frac32\rangle = \frac{i}{\sqrt{2}}(|x;\uparrow  \rangle+i|y;\uparrow  \rangle)  \nonumber\\
|LH_+\rangle &=& |\frac32 , +\frac12\rangle = \frac{i}{\sqrt{6}}(|x;\downarrow  \rangle+i|y;\downarrow  \rangle-2|z;\uparrow \rangle)   \nonumber\\
|LH_-\rangle &=& |\frac32 , -\frac12\rangle = \frac{i}{\sqrt{6}}(|x;\uparrow  \rangle-i|y;\uparrow  \rangle+2|z;\downarrow \rangle) \nonumber\\
|HH_-\rangle &=& |\frac32 , -\frac32\rangle = - \frac{i}{\sqrt{2}}(|x;\downarrow  \rangle-i|y;\downarrow  \rangle)  \nonumber \\
|SO_+\rangle &=& |\frac12 , +\frac12\rangle =\frac{i}{\sqrt{3}}(|x;\downarrow  \rangle+i|y;\downarrow  \rangle+|z;\uparrow \rangle) \nonumber \\
|SO_-\rangle &=& |\frac12 , -\frac12\rangle =\frac{i}{\sqrt{3}}(|x;\uparrow  \rangle-i|y;\uparrow  \rangle-|z;\downarrow \rangle)
\end{eqnarray}
where CB, HH, LH and SO stand for conduction, heavy hole, light hole and split-off bands respectively, and the sign in the subscripts corresponds to the spin component. So, in this basis the 8-band Luttinger-Kohn {\bf k.p} Hamiltonian for a system with an external magnetic field applied, is given by

\begin{equation}\label{eq:KLHamiltonian}
H_{KL}+H^B=    
\begin{pmatrix}
E_{7-} & 0 & -{\sqrt{\frac{1}2}}P^+ & \sqrt{\frac23}P^z & \sqrt{\frac{1}{6}} P^- & 0 & \sqrt{\frac13}P^z & \sqrt{\frac13}P^-   \\
0 & E_{7-} & 0 & -\sqrt{\frac{1}{6}}P^+ & \sqrt{\frac23}P^z & \sqrt{\frac{1}{2}}P^- & \sqrt{\frac13}P^+ & -\sqrt{\frac13}P^z  \\
-\sqrt{\frac{1}{2}}P^- & 0 & E_{8+}^H & B & C & 0 & \sqrt{\frac{1}{2}}B & \sqrt{2}C  \\
\sqrt{\frac23}P^z &-\sqrt{\frac{1}{6}}P^- & B^{\dagger} & E_{8+}^L & 0 & C & -\sqrt{2}A & -\sqrt{\frac32}B  \\
 \sqrt{\frac{1}{6}} P^+ &\sqrt{\frac23}P^z  & C^{\dagger}& 0 &E_{8+}^L & -B & -\sqrt{\frac32} B^{\dagger} & \sqrt{2}A   \\
 0& \sqrt{\frac{1}{2}}P^+ & 0 &C^{\dagger}  & -B^{\dagger} & E_{8+}^H & -\sqrt{2}C^{\dagger} & \sqrt{\frac{1}{2}} B^{\dagger}  \\
 \sqrt{\frac13}P^z& \sqrt{\frac13}P^-&\sqrt{ \frac{1}{2}}B^{\dagger} & -\sqrt{2}A^{\dagger}&-\sqrt{\frac32} B &-\sqrt{2}C & E_{7+} & 0   \\
 \sqrt{\frac13}P^+& -\sqrt{\frac13}P^z&\sqrt{2}C^{\dagger} &-\sqrt{\frac32}B^{\dagger} & \sqrt{2}A^{\dagger}&\sqrt{\frac{1}{2}} B& 0&    E_{7+}
\end{pmatrix} 
\end{equation}
where the $H^B$ term simply adds the Peierls substitution to the canonical momentum $\vec{k}$. The matrix elements of this Hamiltonian are given by
\begin{eqnarray}
\label{matrix:operators}
E_{7-}&=&E_1+\frac{\hbar^2}{2}\Big[\frac12\Big(k_+\frac{1}{m_0}k_- + k_-\frac{1}{m_0}k_+\Big) + k_z\frac{1}{m_0}k_z\Big] \nonumber \\
E_{8+}^H&=&E_2-\frac{\hbar^2}{2m_0}\Big[\frac12\Big(k_+(\tilde{\gamma}_1+\tilde{\gamma}_2)k_- + k_-(\tilde{\gamma}_1+\tilde{\gamma}_2)k_+\Big) + k_z(\tilde{\gamma}_1-2\tilde{\gamma}_2)k_z\Big]\nonumber \\
E_{8+}^L&=&E_2-\frac{\hbar^2}{2m_0}\Big[\frac12\Big(k_+(\tilde{\gamma}_1-\tilde{\gamma}_2)k_- + k_-(\tilde{\gamma}_1-\tilde{\gamma}_2)k_+\Big) + k_z(\tilde{\gamma}_1+2\tilde{\gamma}_2)k_z\Big]\nonumber \\
C&=&\frac{\hbar^2}{2m_0}\sqrt{3}(k_-\bar{\gamma}k_- + k_+\mu k_+)\nonumber \\
B&=&\frac{\hbar^2}{2m_0}\sqrt{3}(k_-\tilde{\gamma}_3k_z + k_z\tilde{\gamma}_3 k_-)\nonumber \\
E_{7+}&=&E_2-\Delta_{SO}-\frac{\hbar^2}{2m_0}\Big[\frac12\Big(k_+\tilde{\gamma}_1k_- + k_-\tilde{\gamma}_1k_+\Big) + k_z\tilde{\gamma}_1k_z\Big]\nonumber \\
A&=&\frac{\hbar^2}{2m_0}\Big[2k_z\tilde{\gamma}_2k_z-\frac12\Big(k_+\tilde{\gamma}_2k_- + k_-\tilde{\gamma}_2k_+\Big)\Big]\nonumber \\
P^{\pm}&=&Pk_{\pm}\nonumber \\
P^z&=&Pk_z\nonumber \\
P&=&\frac{\hbar}{m_0}\langle S|p_x|iX\rangle\nonumber \\
E_P&=&\Big(\frac{2m_0}{\hbar^2}\Big)P^2.
\end{eqnarray}
where $k_{\pm}=e^{\pm i\varphi} (-i(\partial_{\rho}\pm\frac{i}{\rho}\partial_{\varphi})\pm i \frac{\rho}{2l_B^2})$ and $k_z = -i\partial_z$. The parameters involved in this Hamiltonian are the Luttinger-like parameters $\tilde{\gamma_i}$, the Kane energy $E_P$, the Kane parameter $P$, and the magnetic length $l_B=\sqrt{\hbar/2B}$. The energies $E_1$ and $E_2$ are the bottom of the conduction band and top of the valence band energies respectively, and $\Delta_{SO}$ is the split-off energy. Finally, the parameters $\bar{\gamma}$ and $\mu$ are given by $\bar{\gamma}=(\tilde{\gamma}_2+\tilde{\gamma}_3)/2$ and $\mu=(\tilde{\gamma}_2-\tilde{\gamma}_3)/2$.  

The parameters involved in the Zeeman Hamiltonian are the Bohr magneton $\mu_B$, and the Landé's g-factors given by
\begin{equation}
 g_{CB} = 2 - \frac{2E_G\Delta_{SO}}{3E_G(E_G+\Delta_{SO})}
\end{equation}
for the conduction band, and
\begin{equation}
    g_{VB} = \gamma_3 + \frac23\gamma_2 - \frac13\gamma_1 - \frac23
\end{equation}
for the valence band, where $\gamma_i$ are the Lüttinger parameters and $E_G$ the bandgap energy. The parameter $\kappa$ is equal to $\pm1$ for electrons, $\pm3$ for heavy holes, $\pm1$ for light holes and $\pm2$ for the split-off band. The sign of $\kappa$ corresponds to the spin orientation in each band of the Hamiltonian (see Equation~\ref{eq:kpbasis}).
\begin{table}[h!]
\begin{center}
 \begin{tabular}{||p{2cm} p{3cm} p{3cm}  p{4cm}||} 
 \hline
 & GaAs & Ga$_{0.51}$In$_{0.49}$P & (Al$_{0.5}$Ga$_{0.5}$)$_{0.51}$In$_{0.49}$P \\ [0.5ex] 
\hline\hline
E$_G$ (eV)         & 1.519   & 1.9971  & 2.3321 \\ 
\hline
E$_P$ (eV)         & 23.81   & 15.9335 & 14.1007 \\ 
\hline
$\Delta_{SO}$ (eV) & 0.341   & 0.0898  & 0.111 \\ 
\hline
$\gamma_1$         & 7.05    & 4.6296  & 4.4296 \\ 
\hline
$\gamma_2$          & 2.35    & 1.0562  & 1.0945 \\
\hline
$\gamma_3$          & 3       & 1.9997  & 1.933 \\
\hline
$\tilde{\gamma}_1$  & 1.8251  & 1.9701  & 2.4141 \\
\hline
$\tilde{\gamma}_2$  & -0.2625 & -0.2736 & 0.0868 \\
\hline
$\tilde{\gamma}_3$  & 0.3875  & 0.67    & 0.9252 \\ [1ex] 
\hline
\end{tabular}
\caption{Band, Lüttinger and Lütinger-like parameters. \cite{Kishore2014,Saidi2010}}
\label{tabla1}
\end{center}
\end{table}

\begin{table}[h!]
\begin{center}
\begin{tabular}{|| p{5cm} p{3cm}  p{3cm}||} 
\hline
                                                 & CBO (eV) & VBO (eV) \\ [0.5ex] 
\hline\hline
GaAs/Ga$_{0.51}$In$_{0.49}$P                     & 0.1698   & 0.3083    \\ 
\hline
GaAs/(Al$_{0.5}$Ga$_{0.5}$)$_{0.51}$In$_{0.49}$P & 0.385    & 0.4282     \\ [1ex] 
\hline
\end{tabular}
\caption{Band offset parameter for the heterostructures used.  \cite{Kishore2014}}
\label{tabla2}
\end{center}
\end{table}
The parameters for the materials used in this work are given in Tables~\ref{tabla1} and \ref{tabla2}. In all cases, the zero in energy was taken equal to the top of the valence band of GaAs, so the $E_1$ and $E_2$ energies for the other materials can be obtained using the energy offsets in Table~\ref{tabla2}.

\section{Matrix elements}
\label{ap:matrix-elements}
In this Appendix we present some auxiliary calculations that help obtain the matrix elements of the Hamiltonian. If we take the parameter $a=a(\rho,z)$ to be any of the parameters involved in the matrix elements, we can expand some of the expressions given in \ref{matrix:operators} and obtain:

\begin{equation}
\label{eq:kzkz}
 k_z a k_z = -(\partial_z a) \partial_z - a\partial_z^2,
\end{equation}

\begin{equation}
 k_+ak_-+k_-ak_+=2\Big\{\Big(\frac{\rho}{2l_B^2}\Big)^2a - 
\frac{i}{l_B^2}a\partial _\varphi - (\partial_{\rho}a)\partial_{\rho} - 
a(\partial_{\rho}^2+\frac{1}{\rho}\partial_{\rho} 
+\frac{1}{\rho^2}\partial_{\varphi}^2)\Big\},
\end{equation}

\begin{eqnarray}
 k_\pm a k_\pm &=&e^{\pm2i\varphi}\Big\{-\Big(\frac{\rho}{2l_B^2}\Big)^2a\pm 
\frac{\rho}{l_B^2}a\partial_\rho+ 
\frac{i}{l_B^2}a\partial_\varphi \nonumber \\
&&\pm (\partial_\rho a)\frac{\rho}{2l_B^2}- 
(\partial_\rho a)\partial_\rho \mp 
i(\partial_\rho a)\frac{1}{\rho}\partial_\varphi \nonumber \\
&&-a[\partial_\rho^2\mp\frac{2i}{\rho^2}\partial_\varphi
\pm \frac{2i}{\rho}\partial_\varphi\partial_\rho
-\frac{1}{\rho}\partial_\rho -\frac{1}{\rho^2}\partial_\varphi^2]\Big\}
\end{eqnarray}
and
\begin{eqnarray}
\label{eq:k-kz}
 k_-ak_z+k_zak_-=e^{-i\varphi}&&\Big\{\frac{\rho}{l_B^2}a\partial_z - 
(\partial_\rho a)\partial_z - 
2a\partial_z\big(\partial_\rho-\frac{i}{\rho}\partial\varphi\big)-\nonumber \\
&&(\partial_za)\big(\partial_\rho-\frac{i}{\rho}\partial_\varphi+\frac{\rho}{
2l_B^2 } \big)\Big\}.
\end{eqnarray}

In this expressions, when we write $(\partial_i a)$ in parenthesis it means that the only quantity affected by the derivative is the parameter $a$, and not the expressions after it.

When calculating the matrix elements of the Hamiltonian in a basis $\Phi_{\alpha}$, where $\alpha$ are the quantum numbers of the states, it is convenient to write all the material-dependent parameters as functions of the coordinates as:
\begin{eqnarray}
a(\rho,z)&=&a_c[1-\theta(\rho-R_c)][1-\theta(|z|-z_1)] \nonumber \\
        &+& a_s \, [\theta(|z|-z_1) \,\{\theta(|z|-z_2)-\theta(\rho-R_c)\}\,+ \theta(\rho-R_c)] \nonumber \\
        &+&a_b[1-\theta(|z|-z_2)]\theta(|z|-z_1),
\end{eqnarray}
where $a_c$, $a_s$ and $a_b$ is the value of the parameter for the core, shell and barrier materials respectively. Then, the integrals involved are to be calculated as 
\begin{eqnarray}
\label{eq:a}
\langle  a\mathcal{O}\rangle  &=&a_c \int_0^{2\pi}d\varphi \int_{-z_1}^{z_1}dz\int_{0}^{R_c}\rho \,d\rho \,\Phi_{\alpha}^*\mathcal{O}\Phi_{\beta} \nonumber \\
&+&a_s\int_0^{2\pi}d\varphi \int_{-z_1}^{z_1}dz\int_{R_c}^{R}\rho \,d\rho\,\Phi_{\alpha}^*\mathcal{O}\Phi_{\beta} \nonumber \\
&+&a_s \int_0^{2\pi}d\varphi\int_{-L/2}^{-z_2}dz \int_{0}^R \rho d\rho\,\Phi_{\alpha}^*\mathcal{O}\Phi_{\beta} \nonumber \\
&+&a_s \int_0^{2\pi}d\varphi\int_{z_2}^{L/2}dz \int_{0}^R \rho d\rho\,\Phi_{\alpha}^*\mathcal{O}\Phi_{\beta} \nonumber \\
&+&a_b\int_0^{2\pi}d\varphi \int_{-z_2}^{-z_1}dz\int_{0}^{R}\rho \,d\rho\Phi_{\alpha}^*\mathcal{O}\Phi_{\beta} \nonumber \\
&+& a_b\int_0^{2\pi}d\varphi\int_{z_1}^{z_2}dz\int_{0}^{R}\rho \,d\rho \,\Phi_{\alpha}^*\mathcal{O}\Phi_{\beta}
\end{eqnarray}
Here $\mathcal{O}$ is any operator involved in the matrix elements of $H$. Note that if $a(\rho,z)=cte$, as in the model with a one-material nanowire, or in the $E_{7-}$ matrix element, the integrals above come together as a single integral over the whole volume, so if we have for instance $\mathcal{O}=\rho^2$ and $a(\rho,z)=1/m_0$ the integral will be 
\begin{equation}
    \langle a\rho^2 \rangle = \frac{1}{m_0}\int_{0}^{2\pi}d\varphi\int_{-L/2}^{L/2}dz\int_{0}^R\rho d\rho\,\Phi_{\alpha}^*\Phi_{\beta}\rho^2
\end{equation}


When the derivative of the discontinuous parameters, $a(\rho,z)$, enters into the calculation of matrix elements, it results in the appearance of Dirac's deltas and Heaviside step functions on the integrals. Having this in mind, we get
\begin{eqnarray}
\langle (\partial_z a)\, \mathcal{O}\rangle = a_c \int_0^{2\pi}d\varphi\int_0^{R_c}\rho d\rho&&\,\Big[ \Phi_{\alpha}^*(\rho,-z_1,\varphi)\mathcal{O}\Phi_{\beta}(\rho,-z_1,\varphi) -\nonumber \\
&&\Phi_{\alpha}^*(\rho,z_1,\varphi)\mathcal{O}\Phi_{\beta}(\rho,z_1,\varphi)\Big] \nonumber \\
+ a_s \int_0^{2\pi}d\varphi\int_{R_c}^{R}\rho d\rho&&\,\Big[ \Phi_{\alpha}^*(\rho,-z_1,\varphi)\mathcal{O}\Phi_{\beta}(\rho,-z_1,\varphi) -\nonumber \\
&&\Phi_{\alpha}^*(\rho,z_1,\varphi)\mathcal{O}\Phi_{\beta}(\rho,z_1,\varphi)\Big] \nonumber \\
+(a_b-a_s) \int_0^{2\pi}d\varphi\int_0^{R}\rho d\rho&&\,\Big[ \Phi_{\alpha}^*(\rho,-z_2,\varphi)\mathcal{O}\Phi_{\beta}(\rho,-z_2,\varphi) -\nonumber \\
&&\Phi_{\alpha}^*(\rho,z_2,\varphi)\mathcal{O}\Phi_{\beta}(\rho,z_2,\varphi)\Big] \nonumber \\
-a_b \int_0^{2\pi}d\varphi\int_0^{R}\rho d\rho&&\,\Big[ \Phi_{\alpha}^*(\rho,-z_1,\varphi)\mathcal{O}\Phi_{\beta}(\rho,-z_1,\varphi) -\nonumber \\
&&\Phi_{\alpha}^*(\rho,z_1,\varphi)\mathcal{O}\Phi_{\beta}(\rho,z_1,\varphi)\Big] 
\end{eqnarray}
and
\begin{eqnarray}
\label{eq:dra}
\langle (\partial_{\rho}a)\, \mathcal{O}\rangle&=&(a_s-a_c)\int_0^{2\pi}d\varphi\int_{-z_1}^{z_1}dz\,\Phi_{\alpha}^*(R_c,z,\varphi)\mathcal{O}\Phi_{\beta}(R_c,z,\varphi)R_c
\end{eqnarray}

From Equations~\ref{eq:kzkz}-\ref{eq:k-kz} and \ref{eq:a}-\ref{eq:dra}, the calculation of the matrix elements is straightforward. For the functions in Equation~\ref{eq:basis}, the values of the matrix elements can be numerically obtained, although some of the integrals involved can be obtained analytically using the recursion relationships of the Bessel functions, the parity properties of trigonometric functions, and so on \cite{Abramowitz}, thus reducing the numerical error and calculation time.


\begin{thebibliography}{99}

\bibitem{Kishore2014}V. V. Ravi Kishore, B. Partoens and F. M. Peeters, Electronic and optical properties of core–shell nanowires in a magnetic field, J. Phys.: Condens. Matter 26  (2014) 095501. https://doi.org/10.1088/0953-8984/26/9/095501.


\bibitem{Kishore2012}V. V. Ravi Kishore, B. Partoens and F. M. Peeters, Electronic structure of InAs/GaSb core-shell nanowires, Phys. Rev. B 86 (2012) 165439. https://doi.org/10.1103/PhysRevB.86.165439.

\bibitem{Ramos2014}A. Y. Ramos and O. Osenda, Resonance states in a cylindrical quantum dot with an external magnetic field, J. Phys. B: At. Mol. Opt. Phys. 47 (2014) 015502. https://doi.org/10.1088/0953-4075/47/1/015502.

\bibitem{Garagiola2018}M. Garagiola, F. M. Pont and O. Osenda, Excitonic states in spherical layered quantum dots, J. Phys. B: At. Mol. Opt. Phys. 51 (2018) 075504. https://doi.org/10.1016/j.physe.2019.113755.

\bibitem{Avron1977}J. Avron, I. Herbst, and B. Simon, Formation of Negative Ions in Magnetic Fields, Phys. Rev. Lett. 39 (1977) 1068. https://doi.org/10.1103/PhysRevLett.39.1068.

\bibitem{Varga2014} J. A. Salas and K. Varga, He$^-$ in a magnetic field: Structure and stability, Phys. Rev. A 89 (2014) 052501. https://doi.org/10.1103/PhysRevA.89.052501.

\bibitem{Ho1997}Y. K. Ho, Magnetic-field effects on the $^1$P$^o$ shape resonance of H$^-$ above the H(N = 2) threshold, Phys. Lett. A 230 (1997) 190. https://doi.org/10.1016/S0375-9601(97)00252-1.

\bibitem{Sikorski1989}Ch. Sikorski and U. Merkt, Spectroscopy of electronic states in InSb quantum dots,
 Phys. Rev. Lett. 62 (1989) 2164. https://doi.org/10.1103/PhysRevLett.62.2164.

\bibitem{Reed1988}M. A. Reed, J. N. Randall, R. J. Aggarwal, R. J. Matyi, T.
M. Moore, and A. E. Wetsel, Observation of discrete electronic states in a zero-dimensional semiconductor nanostructure, Phys. Rev. Lett. 60 (1988) 535. https://doi.org/10.1103/PhysRevLett.60.535.

\bibitem{Buczko1996}R. Buzcko and F. Bassani, Bound and resonant electron states in quantum dots: The optical spectrum, Phys. Rev. B 54 (1996) 2667. https://doi.org/10.1103/PhysRevB.54.2667.

\bibitem{Bylicki1999}M. Bylicki and W. Jaskolski, Binding of resonant states in a magnetic field, 
Phys. Rev. B 60 (1999) 15924. https://doi.org/10.1103/PhysRevB.60.15924.

\bibitem{Bylicki2005}M. Bylicki, W. Jaskolski, and A. Stachow, Resonance states of two-electron quantum dots, Phys. Rev. B 72 (2005) 075434. https://doi.org/10.1103/PhysRevB.72.075434.

\bibitem{Chakraborty2011} S. Chakraborty and Y. K. Ho, Autoionization resonance states of two-electron atomic systems with finite spherical confinement, Phys. Rev. A 84 (2011) 032515. https://doi.org/10.1103/PhysRevA.84.032515.

\bibitem{Kuros2015} A. Kuro\'s and A. Okopi\'nska, Entanglement Properties of the Two-Electron Quasi-One Dimensional Gaussian Quantum Dot, Few-Body Syst. 56 (2015) 853-858. https://doi.org/10.1007/S00601-015-0992-X.

\bibitem{Sajeev2008} Y. Sajeev and N. Moiseyev, Theory of autoionization and photoionization in two-electron spherical quantum dots, Phys. Rev. B 78 (2008) 075316. https://doi.org/10.1103/PhysRevB.78.075316.

\bibitem{Genkin2010} M. Genkin and E. Lindroth, Effects of screened Coulomb impurities on autoionizing two-electron resonances in spherical quantum dots, Phys. Rev. B 81 (2010) 125315. https://doi.org/10.1103/PhysRevB.81.125315.

\bibitem{Moiseyev1998}N. Moiseyev, Quantum theory of resonances: calculating energies, widths and cross-sections by complex scaling, Phys. Rep. 302 (1998) 211. https://doi.org/10.1016/S0370-1573(98)00002-7.

\bibitem{Sajeev2009} Y. Sajeev, V. Vysotskiy, L. S. Cederbaum, and N. Moiseyev, Continuum remover-complex absorbing potential: Efficient removal of the nonphysical stabilization points, J. Chem. Phys. 131 (2009) 211102. https://doi.org/10.1063/1.3271350.

\bibitem{tomic}S. Tomić, A. G. Sunderland and I. J. Bush, Parallel multi-band k·p code for electronic structure of zinc blend semiconductor quantum dots, J. of Mater. Chem. 16 (2006) 1963-1972. https://doi.org/10.1039/B600701P.

\bibitem{Novik2005}E. G. Novik, A. Pfeuffer-Jeschke, T. Jungwirth, V. Latussek, C. R. Becker, G. Landwehr, H. Buhmann, and L. W. Molenkamp, Band structure of semimagnetic Hg$_{1-y}$Mn$_y$Te quantum wells, Phys. Rev. B 72 (2005) 035321. https://doi.org/10.1103/PhysRevB.72.035321.

\bibitem{Luttinger1955} J. M. Luttinger and W. Kohn, Motion of Electrons and Holes in Perturbed Periodic Fields, Phys. Rev. 97 (1955) 869. https://doi.org/10.1103/PhysRev.97.869.

\bibitem{Hofstadter1976}Douglas R. Hofstadter, Energy levels and wave functions of Bloch electrons in rational and irrational magnetic fields, Phys. Rev. B 14 (1976) 2239. https://doi.org/10.1103/PhysRevB.14.2239

\bibitem{Saidi2010}I. Saïdi, S. Ben Radhia and K. Boujdaria, Band parameters of GaAs, InAs, InP, and InSb in the 40-band k.p model, J. Appl. Phys. 107 (2010) 043701. https://doi.org/10.1063/1.3295900.


\bibitem{Ferron2009}A. Ferrón, O. Osenda, and P. Serra, Entanglement in resonances of two-electron quantum dots, Phys. Rev. A 79 (2009) 032509. https://doi.org/10.1103/PhysRevA.79.032509.

\bibitem{Pont2010}F. M. Pont, O. Osenda, J. H. Toloza, and P. Serra, Entropy, fidelity, and double orthogonality for resonance states in two-electron quantum dots, Phys. Rev. A 81 (2010) 042518. https://doi.org/10.1103/PhysRevA.81.042518.

\bibitem{Krishtopenko2018}S. S. Krishtopenko and F. Teppe, Realistic picture of helical edge states in HgTe quantum wells, Phys. Rev. B 97 (2018) 165408. https://doi.org/10.1103/PhysRevB.97.165408.

\bibitem{Nesbet}R. Nesbet, Variational Principles and Methods in Theoretical Physics and Chemistry, Cambridge University Press, (2004).

\bibitem{Konig2007}Markus König, Steffen Wiedmann, Christoph Brüne, Andreas Roth, Hartmut Buhmann,
Laurens W. Molenkamp, Xiao-Liang Qi and Shou-Cheng Zhang, Quantum spin hall insulator state in HgTe quantum wells, Science 318 (2007) 766. https://doi.org/10.1126/science.1148047.

\bibitem{Zhou2008}Bin Zhou, Hai-Zhou Lu, Rui-Lin Chu, Shun-Qing Shen, and Qian Niu, Finite Size Effects on Helical Edge States in a Quantum Spin-Hall System, Phys. Rev. Lett. 101 (2008) 246807. https://doi.org/10.1103/PhysRevLett.101.246807.

\bibitem{Durnev2016}M. V. Durnev and S. A. Tarasenko, Magnetic field effects on edge and bulk states in topological insulators based on HgTe/CdHgTe quantum wells with strong natural interface inversion asymmetry, Phys. Rev. B 93  (2016) 075434. https://doi.org/10.1103/PhysRevB.93.075434.

\bibitem{Raichev2012}O. E. Raichev, Effective Hamiltonian, energy spectrum, and phase transition induced by in-plane magnetic field in symmetric HgTe quantum wells, Phys. Rev. B 85 (2012) 045310. https://doi.org/10.1103/PhysRevB.85.045310.



\bibitem{Abramowitz}M. Abramowitz and I. Stegun, Handbook of Mathematical Functions with Formulas, Graphs, and Mathematical Tables, ninth Dover printing, tenth GPO printing edition, Dover, New York, 1964.





\end{thebibliography}
\end{document}